# Advanced Reversible Data Hiding With Encrypted Data

Shilpa Sreekumar[1], Vincy Salam[2]

[1] *PG Scholar*, [2] *Assistant Professor*
[1,2] *Department of Computer Science and Engineering*
*Musaliar College of Engineering and Technology, Kerala, India.*

*Abstract*—The advanced RDH work focuses on both data encryption and image encryption which makes it more secure and free of errors. All previous methods embed data without encrypting the data which may subject to errors on the data extraction or image recovery. The proposed work provides a novel RDH scheme in which both data and image can be encrypted and extracted reversibly without any errors. In the proposed work, data extraction and image recovery are free of any errors. The PSNR is significantly improved in the proposed work. This advanced work also performs data hiding in videos.

## I. INTRODUCTION

Reversible data hiding (RDH) in images is a technique, by which the original cover can be recovered after the embedded message is extracted. By first extracting compressible features of original cover and then compressing them efficiently, spare space can be saved for embedding secret data. Many RDH techniques have emerged in recent years like the histogram modification technique and the difference expansion technique. A more popular method is based on difference expansion (DE), in which the difference of each pixel group is expanded, e.g., multiplied by 2, and thus the least significant bits (LSBs) of the difference are all-zero and can be used for embedding messages. Another popular strategy for RDH is histogram shift (HS), in which space is saved for data embedding by shifting the bins of histogram of gray values. The existing scenario focuses on data hiding rather than data encryption and image encryption in the reversible data hiding systems.

In order to provide confidentiality for images, encryption techniques can be employed. Although few RDH methods in encrypted images have been published yet, there are some effective applications if RDH can be applied to encrypted images. In some RDH techniques, the encrypted image is divided into several blocks and by flipping 3 LSBs of the half of pixels in each block, room can be vacated for the embedded bit. The data extraction and image recovery is done by finding which part is flipped in one block. This process can be done with the help of spatial correlation in decrypted image. Some other RDH techniques employed spatial correlations using a different estimation equation and side match method to achieve lower error rate.

In the proposed work, we propose a novel method for RDH in encrypted images with encrypted data. In the proposed work, we first find out space by embedding LSBs of some pixels into other pixels with a traditional RDH method and then encrypt the image, so the positions of these LSBs in the encrypted image can be used to embed data. The encryption of image is realized by using Blowfish encryption algorithm and the secret data is encrypted using Advanced Encryption Standard (AES) algorithm.

## II. PROPOSED WORK

In this advanced reversible data hiding method, encrypted data can be embedded and extracted from both encrypted images and videos. The data is encrypted using AES algorithm and image is encrypted using the Blowfish algorithm. The proposed work also implements digital video watermarking. Video has become an important tool for the entertainment and educational industry. Digital video watermarking is new technology used for copyright protection of digital media. It inserts authentication information in multimedia data which can be used as proof of ownership. Video watermarking algorithms normally prefers robustness. Most of the proposed video watermarking schemes are based on the techniques of image watermarking. The proposed work includes: generation of encrypted data, generation of encrypted image, data embedding, data extraction and image recovery.

### A. Generation of Encrypted data.

The secret data is encrypted using the AES algorithm. First the secret data is encoded using Huffman Encoding before performing AES encryption. Huffman encoding is performed to compress the secret information and then this information is encrypted using AES algorithm. In this processing step, two main algorithms are used: Huffman Encoding and AES algorithm. Huffman's scheme uses a table of frequency of occurrence for each symbol in the input. This table may be derived from the input itself or from data which is representative of the input. AES is based on a design principle known as a substitution permutation network, combination of both substitution and combination, and is fast in both software and hardware.

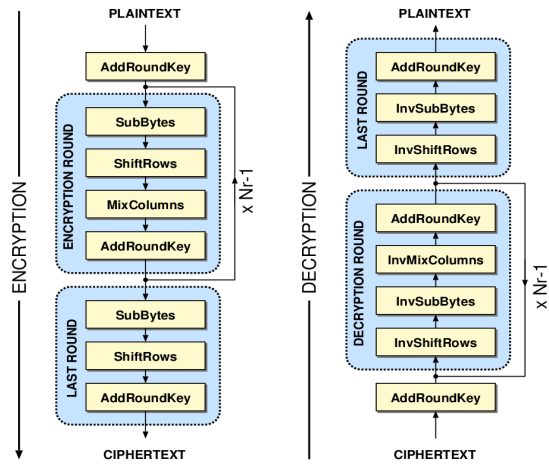

Fig. 1 AES Encryption and Decryption

The key size used for an AES cipher specifies the number of repetitions of transformation rounds that convert the input, called the plaintext, into the final output, called the ciphertext. The proposed work utilizes the 128-bit key size of the AES algorithm. Each round consists of four processing steps in which the first step is the substitute byte step and next is the shift row transformation, third is the mix column transformation and last step is the addroundkey transformation step. A set of reverse rounds are applied to transform ciphertext back into the original plaintext using the same encryption key.

1  Algorithm: AES Cipher

**Cipher**(bytein[16], byteout[16], key_array round_key[Nr+1])

   **begin**

      byte state[16];
      state = in;
      AddRoundKey(state, round_key[0]);
      **for** i = 1 to Nr-1 stepsize 1 **do**
      SubBytes(state);
      ShiftRows(state);
MixColumns(state);
      AddRoundKey(state, round_key[i]);
   **end for**
      SubBytes(state);
      ShiftRows(state);
      AddRoundKey(state, round_key[Nr]);

**end**

*B. Generation of Encrypted image.*

The next step after data encryption is image encryption which is done using Blowfish algorithm. Blowfish is a 64-bit symmetric block cipher that uses a variable-length key from 32 to 448-bits (14 bytes). The algorithm was developed to encrypt 64-bits of plaintext into 64-bits of cipher text efficiently and securely. The operations selected for the algorithm were table lookup, modulus, addition and bitwise exclusive-or to minimize the time required to encrypt and decrypt data on 32-bit processors. Blowfish incorporates a 16 round Feistel network for encryption and decryption. But during each round of Blowfish, the left and right 32-bits of data are modified unlike DES which only modifies the right 32-bits to become the next round's left 32-bits. Blowfish incorporated a bitwise exclusive-or operation to be performed on the left 32-bits before being modified by the F function or propagated to the right 32-bits for the next round. Blowfish also incorporated two exclusive-or operations to be performed after the 16 rounds and a swap operation. This operation is different from the permutation function performed in DES. The diagram of action of Blowfish is shown in figure 2.

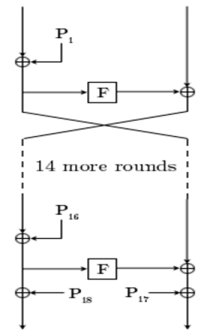

Fig. 2 Feistel structure of Blowfish

In Figure 2, each line represents 32 bits. The algorithm keeps two subkey arrays: the 18-entry P-array and four 256-entry S-boxes. The S-boxes accept 8-bit input and produce 32-bit output. One entry of the P-array is used every round, and after the final round, each half of the data block is XOR with one of the two remaining unused P-entries.

1. Algorithm: Blowfish Encryption

   **Divide** x into two 32-bit halves: xL, xR

   **For** i = 1 to 16:

      xL = XL XOR Pi

      xR = F(XL) XOR xR

      **swap** XL and xR

      **swap** XL and xR

      xR = xR XOR P17

      xL = xL XOR P18

   **Recombine** xL and xR.

*C. Data hiding in Encrypted image.*

After image encryption, the encrypted secret data is embedded into the encrypted image by employing a traditional RDH algorithm like Histogram modification method or a LSB replacement method. Here data embedding is performed in color images. Here each pixel in color images will have three individual components Red(R), Green(G) and Blue(B). The pixel values of these color components will be in the range of [0 255]. The message bits can be embedded in all the three planes and these planes can be recombined to form the original color image. Here the message bits are embedded in every Red component in the RGB plane. After the data embedding is done, the PSNR value is calculated and shown in the textbox in the MATLAB simulator.The proposed work also performs data hiding in videos which can be used for copyright protection of digital media. Here video is divided into frames and this RGB frames are converted to YUV frames. Frames are sequence of high resolution images and the data embedding is performed by looping of frames.

*D. Data Extraction and Image Recovery*

After the data embedding process, the embedded image is obtained along with the PSNR value. The next step is data extraction process which is the reverse of the data embedding process. Here encrypted data is extracted from the encrypted image in the reverse order by employing the AES Decryption algorithm. After that the original image is extracted by using Blowfish Decryption algorithm. After performing the AES Decryption, the Huffman encoded data is retrieved and then Huffman decoding is performed to retrieve the original data. This same process is applied to videos and data extraction and image recovery is successfully separated in videos using the AES algorithm and Blowfish algorithm.

III   PERFORMANCE EVALUATION AND RESULTS

The proposed method can achieve real reversibility as compared to previous existing methods and the PSNR value is significantly improved using this novel method. The PSNR is calculated using the formula:

$$\text{PSNR(dB)} = 10 * \log\left(\frac{255^2}{MSE}\right)$$

$$\text{MSE} = \sum_{i=1}^{x}\sum_{j=1}^{y}\frac{|(A_{ij}-B_{ij})|^2}{x*y}$$

Where   MSE:  Mean-Square Error
　　　　x:    width of the image
　　　　y:    height of the image
　　　　x * y: number of pixels.

The simulation result is as shown below:

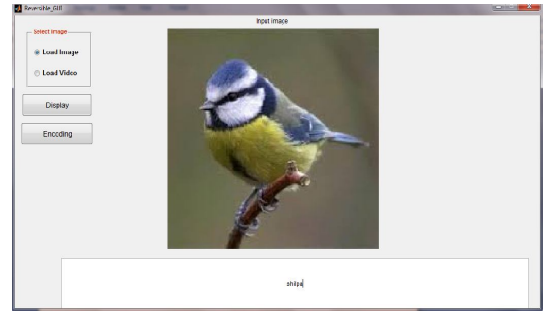

Fig.3  Secret data is entered into the textbox given below the input image.

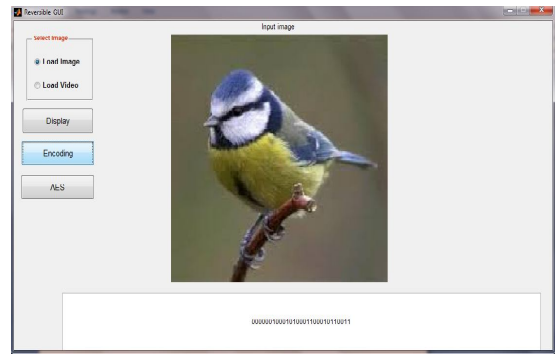

Fig.4  Secret data is encoded using Huffman Encoding

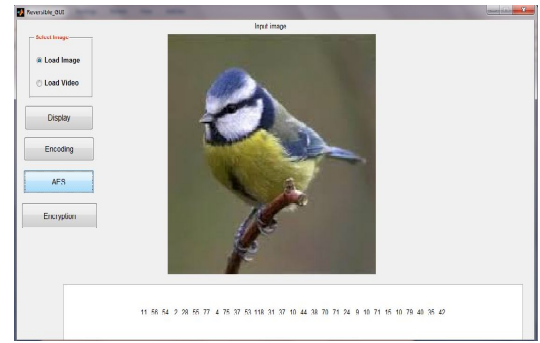

Fig.5  AES Encryption is performed on encoded data.

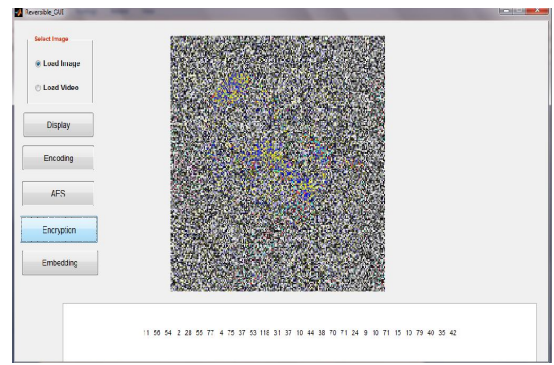

Fig.6   Image Encryption is employed using Blowfish**.**

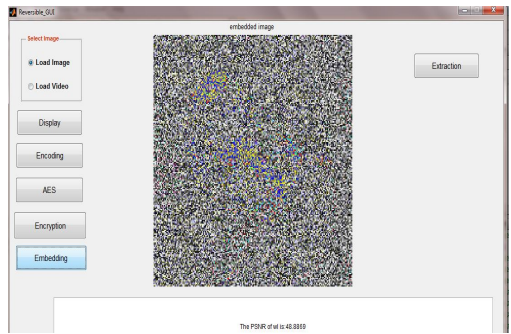

Fig.7 Encrypted data is embedded into the encrypted image and the PSNR is 48.88dB.

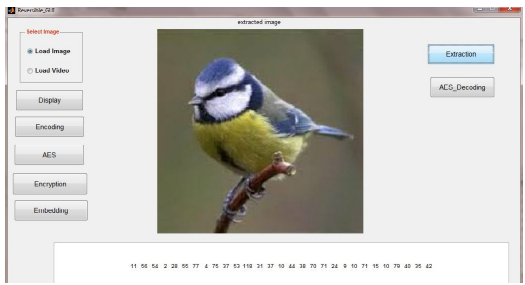

Fig.8 Original Image is extracted and encrypted data is obtained.

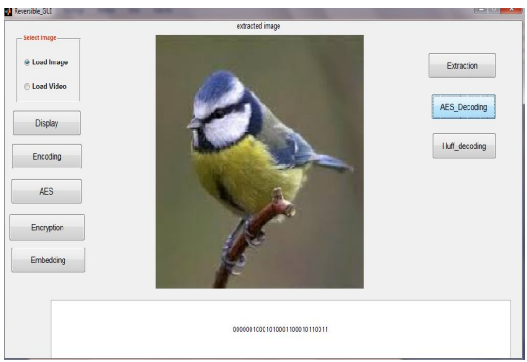

Fig.9 AES decryption

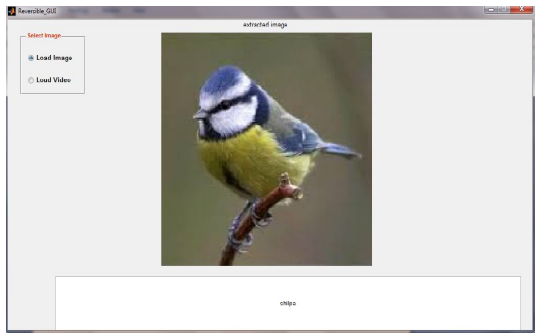

Fig. 10 Original secret data obtained after Huffman decoding.

### III. CONCLUSIONS

An advanced RDH scheme with encrypted data has been presented in this paper. This work combines data encryption with image encryption. The two main algorithms implemented for data encryption and images encryption are the Advanced Encryption Standard (AES) algorithm and the Blowfish algorithm. The work begins with data encoding step which is performed by employing Huffman encoding method and this is done to compress the data. The next step is data encryption which is performed using AES algorithm and after this step the image is encrypted using the Blowfish algorithm which is highly secure because of its longer key length and strongest and fastest nature in data processing compared to other algorithms. Apart from data hiding in images, the proposed work can also performs data hiding in videos which takes this work to a new level in the advanced RDH scheme.

### REFERENCES

[1] Kede Ma, Weiming Zhang, Xianfeng Zhao, Member, IEEE, NenghaiYu, and Fenghua Li"Reversible Data Hiding in Encrypted Images byReserving Room Before Encryption" ieee transactions on information forensics and security, vol. 8, no. 3, march 2013.

[2] M. Goljan, J. Fridrich, and R. Du, "Distortion-free data embedding," in Proc. 4th Int. Workshop on Information Hiding, Lecture Notes in Computer Science, 2001, vol. 2137, pp. 27–41.

[3] M. U. Celik, G. Sharma, A. M. Tekalp, and E. Saber, "Lossless generalized- LSB data embedding," IEEE Trans. Image Process., vol. 14, no. 2, pp. 253–266, Feb. 2005.

[4] J. Fridrich, M. Goljan, and R. Du, "Lossless data embedding for all image formats," in Proc. Security and Watermarking of Multimedia Contents IV, Proc. SPIE, 2002, vol. 4675, pp. 572–583.

[5] J. Tian, "Reversible data embedding using a difference expansion," IEEE Trans. Circuits Syst. Video Technol., vol. 13, no. 8, pp. 890–896, Aug. 2003

[6] A. M. Alattar, "Reversible watermark using the difference expansion of a generalized integer transform," IEEE Trans. Image Process., vol. 13, no. 8, pp. 1147–1156, Aug. 2004.

[7] Ratinder Kaur, V. K. Banga "Image Security using Encryption based Algorithm" International Conference on Trends in Electrical, ElectronicsandPowerEngineering (ICTEEP'2012) July 15-16, 2012 Singapore.

[8] Pia Singh Prof. Karamjeet Singh "Image encryption and decryption using blowfish algorithm in matlab" International Journal of Scientific & Engineering Research, Volume 4,Issue 7, July- 2013 150 ISSN 2229-5518.

[9] Prachi V. Powar , Prof. S.S.Agrawal "Design of digital video watermarking scheme using matlab simulink"PRACHI V POWAR* et al ISSN: 2319 1163 Volume: 2 Issue: 5 826 - 830 IJRET, MAY 2013.